\documentclass[preprint,aps]{revtex4}
\usepackage{amsmath}
\usepackage{amsfonts}
\usepackage{mathrsfs}
\usepackage{graphicx}
\usepackage{times}
\usepackage{color}

\def\be{\begin{equation}}
\def\ee{\end{equation}}
\def\bea{\begin{eqnarray}}
\def\eea{\end{eqnarray}}

\begin{document}

\title{$p$-wave chiral superfluidity from an $s$-wave interacting atomic
Fermi gas}
\author{Bo Liu $^{1}$, Xiaopeng Li$^{1,2,3}$, Biao Wu$^{4,5}$, and W.
Vincent Liu $^{1}$}
\email{w.vincent.liu@gmail.com}
\affiliation{$^{1}$Department of Physics and Astronomy, University of Pittsburgh,
Pittsburgh, PA 15260, USA\\
$^{2}$ Condensed Matter Theory Center, University of Maryland, College Park,
MD 20742, USA\\
$^{3}$Joint Quantum Institute, University of Maryland, College Park, MD
20742, USA\\
$^{4}$International Center for Quantum Materials, Peking University, Beijing
100871, China\\
$^{5}$Collaborative Innovation Center of Quantum Matter, Beijing, China}
\date{\today}

\begin{abstract}
Chiral $p$-wave superfluids are fascinating topological quantum states of
matter that have been found in the liquid $^3$He-A phase and arguably in the
electronic Sr$_2$RuO$_4$ superconductor. They are shown fundamentally
related to the fractional $5/2$ quantum Hall state which supports fractional
exotic excitations. A common understanding is that such states require
spin-triplet pairing of fermions due to $p$-wave interaction. Here we report
by controlled theoretical approximation that a center-of-mass Wannier $p$%
-wave chiral superfluid state can arise from spin-singlet pairing for an $s$%
-wave interacting atomic Fermi gas in an optical lattice. Despite a
conceptually different origin, it shows topological properties similar to
the conventional chiral $p$-wave state. These include a non-zero Chern
number and the appearance of chiral fermionic zero modes bounded to domain
walls. Several signature quantities are calculated for the cold atom
experimental condition.
\end{abstract}

\maketitle

\textit{Introduction.---} Topological superconductors, like the type of $%
p_x+ip_y$-wave pairing studied in the liquid $^3$He~\cite{2003_Volovik_book}
and strontium ruthenates~\cite{2012_Kallin_ReportsProgress}, are among the
most desirable unconventional many-body states in condensed matter physics~%
\cite{2008_Nayak_RMp}. In two dimensions, their topological
properties are fundamentally linked to a class of fractional quantum
Hall states of non-Abelian statistics~\cite{2000_ReadGreen_PRB}.
Studies of vortices in such materials point to fascinating braiding
statistics and applications in topological quantum computing. The
fate of topological superconductivity in two-dimensional electronic
matter remains however debatable. In the field of ultracold atoms,
this phase was predicted to appear near the $p$-wave Feshbach
Resonance in Fermi gases~\cite{2007_Leo_AnnalsPhy}. However, the
life time of such systems is severely limited by the three-body
collisions, and achieving superfluidity in the resonance regime was
found experimentally challenging~\cite{2003_Jin_PRL}. Other
strategies like spin-orbit coupling or dipolar interaction also meet
new difficulties such as heating or ultracold chemical
reactions~\cite{2013_Spielman_nature,2012_Baranov_Reviews}.  There
is a separate approach being proposed to get around---hybridizing
materials of separate topological and superconducting properties,
which also encounters some engineering
difficulties~\cite{2011_Xiaoliang_RevModPhys}. After all, the search
for the homogeneous chiral $p$-wave superconductivity in two
dimensions has stood largely open for both electronic and atomic
matter systems.

Here we report the discovery of a new mechanism to achieve chiral
topological superfluidity. We shall demonstrate this with cold fermionic
atoms in optical lattices with the model to be introduced below. The key
concept dramatically departing from the conventional wisdom that relies on
the $p$ or higher partial wave pairing in relative motion is to keep the
fermion interaction within the usual $s$-wave channel by pairing fermions
from different Wannier orbitals, and the center-of-mass orbital motion of
condensed pairs is examined for possible nontrivial topology. Recently the
research of higher orbital bands in optical lattices has evolved rapidly~%
\cite{2011_Vincent_naturephy}, where the orbital degrees of freedom are
found to play a crucial role as in solid state materials. From the early
experimental attempt~\cite{2007_Bloch_PRL} to the breakthrough observation~%
\cite{2011_Hemmerich_NatPhys} of long-lived $p$-band bosonic atoms in a
checkerboard lattice, a growing evidence points to an exotic $p_x+ip_y $
orbital Bose-Einstein condensate~\cite{2011_Vincent_naturephy}. For fermions
with attractive interaction, superfluid states similar to the type of
Fulde-Ferrell-Larkin-Ovchinnikov were found in the theoretical studies of
pairing in the $p$-bands~\cite{2011_Zicai_PRA} and that between the $s$-band
and a single $p$-band~\cite{2010_Zixu_PhysRevA}. As we shall show with the
model below, pairing fermions from the orbitals of different angular momenta
can lead to other unexpected results.

\begin{figure}[t]
\begin{center}
\includegraphics[width=8cm]{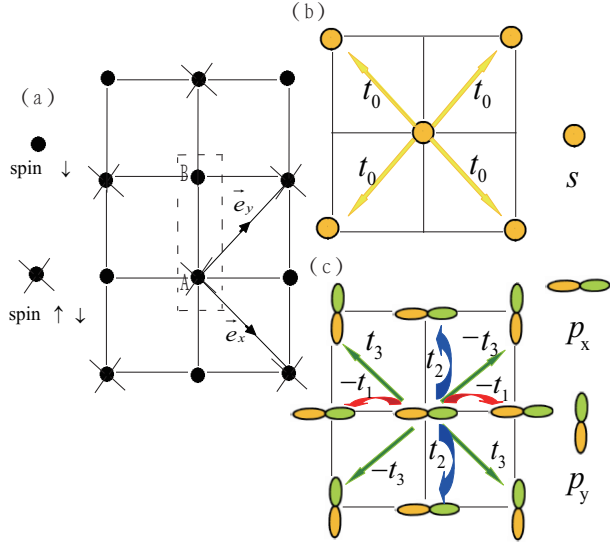}
\end{center}
\caption{(a) Schematic picture of a 2D spin-dependent optical lattice, where
the spin up (s orbital band) and down (p orbital band) component lying
within different geometry lattice potential, respectively. Here A and B
stand for two different sites in one unit cell, $\vec{e}_x$ and $\vec{e}_y$
are the primitive unit vectors; (b) and (c) Schematic views illustrate
tunneling $t_0$, $t_1$, $t_2$ and $t_3$ of fermions prepared in the s and p
orbitals, respectively.}
\label{fig:lattice}
\end{figure}

Let us consider an attractive $s$-wave interacting Fermi gas composed of two
hyperfine states, to be referred to as spin $\uparrow$ and $\downarrow$,
loaded in a spin-dependent 2D optical lattice shown in Fig.~\ref{fig:lattice}%
(a). The spin dependence of the lattice is motivated by various theoretical
designs ~\cite{2004_Vincent_PhysRevA,2011_Daley_QIP} and most importantly
the recent experimental demonstration of it with bosons~\cite%
{2010_DeMarco_NJP,2012_Sengstock_Natphys}. Further let the gas be tuned with
a population imbalance between the two spin species by the techniques
developed in the recent experimental advances~\cite%
{2006_Martin_Science,2006_Hulet,2010_Salomon_nature}. A key condition that
we propose here is to tune the population imbalance (or equivalently the
chemical potential difference) sufficiently large such that the spin $%
\uparrow$ and $\downarrow$ Fermi levels reside in the $s$ and $p$
orbital bands, respectively. The rotation symmetry ($C_4$) of the
lattice dictates that the two $p$ orbital bands, $p_x$ and $p_y$,
are degenerated at the high symmetry points in the momentum space.
Later on we shall see that this symmetry and hence degeneracy are
necessary for the $p_x+ip_y$-wave paired superfluidity. Technically
speaking, the Bravais lattices for the spin up and down fermions are
$45^{\circ}$ rotated from each other.

\textit{Effective model.---}  A system of fermionic atoms, say
$^6$Li, loaded into an optical lattice (Fig.~\ref{fig:lattice}) in
the tight binding regime is described by a multi-orbital Fermi
Hubbard model
\begin{equation}
H=H_{0}+H_{\mathrm{int}},  \label{eq:Ham}
\end{equation}
where $H_0$ describes tunneling pictorially represented in Fig.\ref%
{fig:lattice}(b) and (c) (the expression for $H_{0}$ is  standard and is
given in Supplementary Materials) and $H_{\mathrm{int}}$ is the Hubbard
interaction,
\begin{eqnarray}
H_{\mathrm{int}} &=&-U\sum_{\mathbf{R}}[C_{s}^{A\dagger }(\mathbf{R}%
)C_{s}^{A}(\mathbf{R})-\frac{1}{2}][C_{p_{x}}^{A\dagger }(\mathbf{R}%
)C_{p_{x}}^{A}(\mathbf{R})  \notag \\
&&+C_{p_{y}}^{A\dagger }(\mathbf{R})C_{p_{y}}^{A}(\mathbf{R})-1].
\label{eq:Hint}
\end{eqnarray}%
Here $C_{\nu }^{A}(\mathbf{R})$ and $C_{\nu }^{B}(\mathbf{R})$ are fermionic
annihilation operators for the localized $\nu $ ($s$, $p_{x}$ or $p_{y}$)
orbitals on $A$ and $B$ sites, respectively. The interactions between $s$
and $p$ orbitals originate from interactions between two hyperfine states,
which are tunable by the $s$-wave Feshbach Resonance in ultracold atomic
gases. We focus on the case with attractive interaction where
superconducting pairing is energetically favorable.

The system, as described by the Hamiltonian in Eq.~\eqref{eq:Ham} exhibits
lattice rotation $C_4$ and reflection symmetries. For the reflection in the $%
x$ and $y$ direction, the fermionic operators transform as $\mathcal{R}%
_{x}\equiv\{C^{A\setminus B}_{p_{x}}(\mathbf{R})\rightarrow -C^{A\setminus
B}_{p_{x}}(-R_{y},-R_{x}), C^{A\setminus B}_{p_{y}}(\mathbf{R})\rightarrow
C^{A\setminus B}_{p_{y}}(-R_{y},-R_{x})\}$ and $\mathcal{R}_{y} \equiv \{
C^{A\setminus B}_{p_{x}}(\mathbf{R})\rightarrow C^{A\setminus
B}_{p_{x}}(R_{y},R_{x}), C^{A\setminus B}_{p_{y}}(\mathbf{R})\rightarrow
-C^{A\setminus B}_{p_{y}}(R_{y},R_{x})\}$, respectively. Under the lattice
rotation, $C^{A\setminus B}_{p_{x}}(\mathbf{R})\rightarrow C^{A\setminus
B}_{p_{y}}(-R_{y},R_{x}), C^{A\setminus B}_{p_{y}}(\mathbf{R})\rightarrow
-C^{A\setminus B}_{p_{x}}(-R_{y},R_{x}).$ These symmetries, reflection
symmetries in particular, play an essential role in the following theory.

\textit{Two-Flavor Ginzburg-Landau theory.---}  From the analysis of
Cooper's problem (see Supplementary Materials), we conclude that
condensation of Cooper pairs at $\mathbf{Q}=(\pi/a,\pi/a)$ is
energetically favorable for the ground state, where $a$ is the
lattice constant. Then, it is convenient
to introduce two slowly varying bosonic fields $\Delta_{x}(\mathbf{x})$ and $%
\Delta_{y}(\mathbf{x})$, which represent Cooper pairs $(-1)^{R_{x}+R_{y}}U%
\langle C_{p_{x}}^{A}(\mathbf{R})C_{s}^{A}(\mathbf{R})\rangle $ and $%
(-1)^{R_{x}+R_{y}}U\langle C_{p_{y}}^{A}(\mathbf{R})C_{s}^{A}(\mathbf{R}%
)\rangle $, respectively. That gives a two-flavor Ginzburg-Landau free
energy respecting all the symmetries of the microscopic model as follows
\begin{equation}
F[\Delta _{x},\Delta _{y}]=\int d^{2}\mathbf{x}\left[ f_{\mathrm{Mean}}(%
\mathbf{x})+f_{\mathrm{Gaussian}}(\mathbf{x})\right] ,
\label{eq:Totalfreeenergy}
\end{equation}
with $f_{\mathrm{Mean}} = r(|\Delta _{x}|^{2}+|\Delta
_{y}|^{2})+g_{1}(|\Delta_{x}|^{4}+|\Delta _{y}|^{4}) + g_{2}|\Delta
_{x}|^{2}|\Delta _{y}|^{2}+g_{3}(\Delta _{x}^{\ast }\Delta _{x}^{\ast
}\Delta _{y}\Delta _{y}+h.c.)$, and $f_{\mathrm{Gaussian}} = K(|\partial
_{x}\Delta _{x}|^{2}+|\partial_{y}\Delta_{y}|^{2} + |\partial _{x}\Delta
_{y}|^{2}+|\partial_{y}\Delta _{x}|^{2})$.

This free energy generalizes the theory of two-gap superconductivity as
proposed in the context of transition metals\cite{2002_Babaev_PRL}. We have
neglected temporal fluctuations of Cooper pair fields and such a treatment
is valid at finite temperature away from quantum critical regime. In this
theory, we want to emphasize two key points due to the reflection
symmetries: first, $\Delta _{x}$ and $\Delta _{y}$ are decoupled at
quadratic level; second, linear derivatives such as $\Delta _{x}^{\ast
}\partial _{x}\Delta _{x}+\Delta _{x}^{\ast }\partial _{y}\Delta _{x}$ are
prohibited. The absence of linear derivatives makes the fluctuations of $%
\Delta_{x\setminus y}$ suppressed, and condensation of Cooper pairs at $%
(\pi/a, \pi/a)$ is expected to be stable at least when $t_{3}$ is
infinitesimal. For finite $t_3$ the stability (i.e., $K>0$ in Eq.~%
\eqref{eq:Totalfreeenergy}) is confirmed in our numerics (see Supplementary
Materials).

\begin{figure}[t]
\begin{center}
\includegraphics[width=8cm]{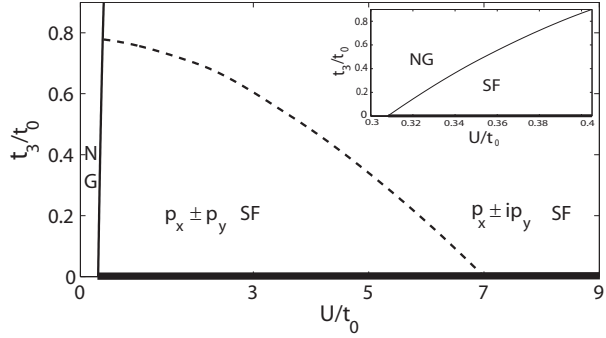}
\end{center}
\caption{Zero-temperature phase diagram--The solid line illustrates the
phase transition from normal gas (NG) to superfluid state. When $U/t_0<7$,
the critical value of $t_3/t_0$ as shown by the dash line, beyond this
threshold a phase transition from $p_x\pm p_y$ to $p_x\pm ip_y$ superfluid
state occurs. When $U/t_0 \geq 7$, $p_x\pm ip_y$ superfluid state is the
ground state with non-zero $t_3$. The thick solid line stands for a
two-component superfluid state.}
\label{fig:phasediag}
\end{figure}

With $r$ and $g_3$ obtained from integrating out fermions, we find a phase
diagram shown in Fig.~\ref{fig:phasediag}. With moderate attraction $U <7t_0$%
, a first order phase transition from the $p_x \pm p_y$ to $p_x \pm ip_y$
phase occurs when $t_3$ is above some critical value. Surprisingly, when the
attraction is strong enough $U>7 t_0$, we find that even infinitesimal $t_3$
makes the $p_x\pm ip_y$ favorable, opening a wide window for this
non-trivial state. When $t_3=0$, the system has $U(1)\times U(1)$ symmetry,
which means no phase coherence between the two components $\Delta_x$ and $%
\Delta_y$. We also study the finite temperature phase transitions (see
Supplementary Materials) and find that the Kosterlitz-Thouless transition
temperature can reach about $109$nk, being accessible in the current
experiments\cite{2008_Bloch_science,2008_Esslinger_nature}, when the lattice
strengths are $V_{s}/E_{R}=3$ and $V_{p}/2E_{R}=5$ for $s$ and $p$ orbitals,
respectively.

\textit{Gapless chiral fermions.---}  We now show that the $p_{x}\pm
ip_{y}$ superfluid state possesses important measurable signatures
due to the broken time reversal $Z_{2}$ symmetry which belongs to
the Ising universality class. Following the standard procedure, our
calculation finds that the state is topologically nontrivial by a
non-zero Chern number, which is $1$ and $-1$ for the $p_{x}+ip_{y}$
and $p_{x}-ip_{y}$ state, respectively. The topological properties
are manifested in the existence of gapless chiral fermions, emergent
on a domain wall connecting topologically distinct regions. In
experiments, Ising domains of $p_{x}+ip_{y}$ and $p_{x}-ip_{y}$ are
expected to spontaneously form as have been observed in the recent
cold atom experiment studying ferromagnetic
transitions~\cite{2013_Chin_Natphys}. In the following, we show that
a domain wall defect carrying gapless fermions as bounded surface
states is experimentally accessible.

\begin{figure}[t]
\begin{center}
\includegraphics[width=10.8cm]{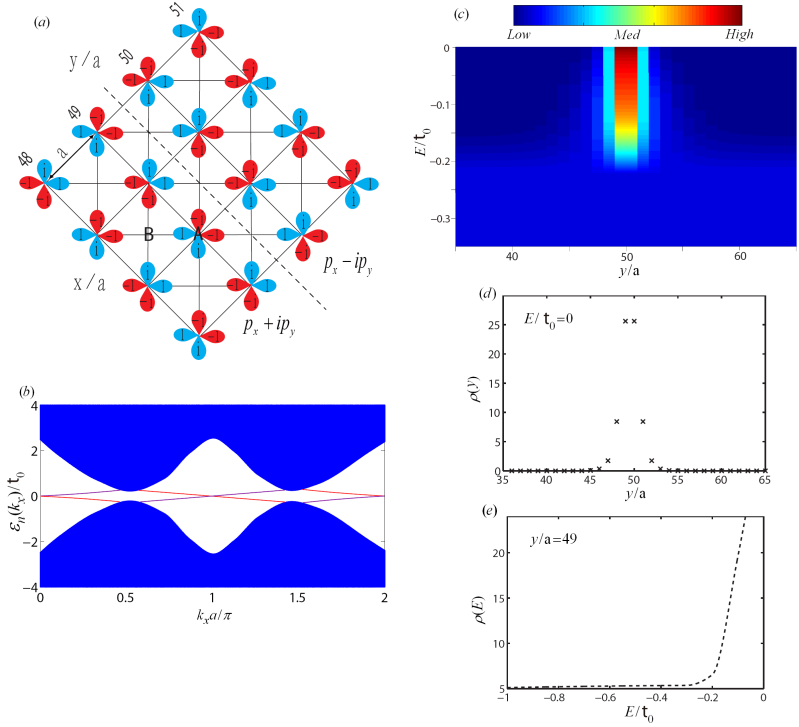}
\end{center}
\caption{(a) Schematic picture of a lattice system in the presence of a
domain wall. (b) Energy spectrum of the system with a domain wall defect,
when $t_1/t_0=8$, $t_2/t_0=2$, $t_3/t_0=0.1$ and $U/t_0=9$. The purple and
red branches correspond to the modes at the domain wall and the edge of the
lattice, respectively. (c), (d), and (e) show the local density of states
(LDOS) defined in the main text. The peak of LDOS located at domain wall is
shown by red color in (c) and further shown with $E/t_0=0$ and $y/a=49$ in
(d) and (e), respectively. The LDOS is in units of $1/at_0$.}
\label{fig:dwenergy}
\end{figure}

Considering a lattice geometry in the presence of a domain wall
decorated superconducting background as in Fig.~\ref{fig:dwenergy}
(a), the mean-field Hamiltonian is given by
\begin{align}
H_{M}& =H_{0}-U\sum_{\mathbf{R}}[C_{s}^{A\dagger }(\mathbf{R}%
)C_{p_{x}}^{A\dagger }(\mathbf{R})<C_{p_{x}}^{A}(\mathbf{R})C_{s}^{A}(%
\mathbf{R})>+<C_{s}^{A\dagger }(\mathbf{R})C_{p_{x}}^{A\dagger }(\mathbf{R}%
)>C_{p_{x}}^{A}(\mathbf{R})C_{s}^{A}(\mathbf{R})  \notag \\
& +C_{s}^{A\dagger }(\mathbf{R})C_{p_{y}}^{A\dagger }(\mathbf{R}%
)<C_{p_{y}}^{A}(\mathbf{R})C_{s}^{A}(\mathbf{R})>+<C_{s}^{A\dagger }(\mathbf{%
R})C_{p_{y}}^{A\dagger }(\mathbf{R})>C_{p_{y}}^{A}(\mathbf{R})C_{s}^{A}(%
\mathbf{R})]  \notag \\
& +U\sum_{\mathbf{R}}\{C_{s}^{A\dagger }(\mathbf{R})C_{s}^{A}(\mathbf{R})+%
\frac{1}{2}[C_{p_{x}}^{A\dagger
}(\mathbf{R})C_{p_{x}}^{A}(\mathbf{R})+C_{p_{y}}^{A\dagger
}(\mathbf{R})C_{p_{y}}^{A}(\mathbf{R})]\}. \label{eq:Hmean}
\end{align}
The energy spectrum of fermionic excitations is obtained by
diagonalizing Eq.~\eqref{eq:Hmean}. With the periodical boundary
condition chosen in the $x$ direction (Fig.~\ref{fig:dwenergy}(a)),
the momentum $k_{x}$ is a good quantum number and the energy spectra
in Fig.~\ref{fig:dwenergy}(b) is thus labeled by $k_{x}$. For the
same reason as in quantum Hall insulators, the number of gapless
chiral modes moving along the interface is topologically determined
by the difference of the Chern numbers in regions on either side of
the interface~\cite{2008_Haldane_PRL}; in this case $|\Delta C|=2$.
This conclusion is confirmed in our numerics. As shown in Fig.~\ref{fig:dwenergy}%
(b), we find four gapless chiral modes, with two localized on the domain
wall (purple color) and the other two on the outer edges of the lattice (red
color). From their spectra $\varepsilon _{n}({k_{x}})$, the two chiral modes
on the domain wall have positive group velocities, which lead to anomalous
mass flow along the domain wall. To characterize the localization of chiral
fermions, we calculate the local density of states (LDOS) $\rho
(y,E)=1/2\sum_{n,\nu }\int dk_{x}[|u_{n}^{\nu }|^{2}\delta (E-\varepsilon
_{n})+|v_{n}^{\nu }|^{2}\delta (E+\varepsilon _{n})],$ where $(u_{n}^{\nu
},v_{n}^{\nu })^{T}$ is the eigenvector corresponding to the eigenenergy $%
\varepsilon _{n}$ of Hamiltonian Eq.~\eqref{eq:Hmean} and $\nu $
runs over all the Wannier orbitals ($s$, $p_{x}$ or $p_{y}$) on $A$
and $B$ sites.
The peak of LDOS located at the position of the domain wall, as shown in Fig.~%
\ref{fig:dwenergy}(c), (d) and (e), illustrates the existence of localized
gapless surface states, reminiscent of the quantum Hall edge states. Taking
a laser wavelength of $\lambda =1024$nm typical for the current optical
lattices, the width of the LDOS peak is estimated about $2\mu m$. This is
greater than the reported spatial resolution (about $1.4\mu m$) in the radio
frequency spectroscopy measurement~\cite{2007_Ketterle_PRL}, which makes the
detection of this signal experimentally accessible.

In summary, when studying a spin imbalanced atomic Fermi gas with an $s$%
-wave interaction, we find surprisingly a topological $p$-wave superfluid
state whose pairing symmetry and topological origin differ from the previous
known superconducting or superfluid phases. To emphasize a remarkable
difference, this phase does not require an interaction beyond the usual
attractive $s$-wave component. Hence a short-ranged contact interaction as
has been widely realized in cold gases should satisfy well. A key concept is
the fermionic Cooper pairing between the orbitals of different angular
momenta in an optical lattice. For the example presented here, they are the
parity even $s$ and odd $p$ orbitals. The $p$-wave symmetry refers to the
center-of-mass motion, not to the relative motion of each fermion pair as in
the well-known $^3$He superfluid. For free or repulsively interacting
systems, previous studies found that mixing orbitals of opposite parities
leads to topological semimetal and insulator phases~\cite%
{2013_xiaopeng_Natcommun,2012_Kaisun_naturephy}. Whether or how the two
phenomena from either sign of the interaction are topologically related is
an intriguing question for the future research.

Experimentally, one may consider the existing proposals for
realizing spin-dependent optical
lattices~\cite{2004_Vincent_PhysRevA,2011_Daley_QIP}. Alternatively,
the recent progress in group-II (alkaline-earth-metal) atoms points
to the possibility of having even greater spin-dependence tunability
if to load two-species fermionic atoms from the ground $^1S_0$ level
and the long-lived metastable levels like $^3P_0$ and to take the
advantage of the atomic orbit dependent AC Stark effect. This should
in principle be able to make the lattices for different components
being completely independent (so maximally spin-dependent lattice)
by selection of the appropriate wavelengths~\cite{2011_Daley_QIP}.
The appearance of chiral fermionic zero modes bounded to domain
walls associated with the orbital Ising order is predicted to be a
fascinating and concrete experimental signature for this novel
state. Both zero and finite temperature phase diagram are also
established, providing the estimates for potential experiments.

\textit{Acknowledgements.---}  The authors want to thank Randy Hulet
and Andrew Daley for helpful discussions. This work is supported by
AFOSR (FA9550-12-1-0079), ARO (W911NF-11-1-0230), DARPA OLE Program
through ARO and the Charles E. Kaufman Foundation of The Pittsburgh
Foundation (B.L., X.L. and W.V.L.), the National Basic Research
Program of China (Grant No. 2013CB921903, 2012CB921300) and NSF of
China (11274024, 11334001) (B.W.), and Overseas Collaboration
Program of NSF of China (11128407) (W.V.L., B.W.). X.L. acknowledges
support by JQI-NSF-PFC, ARO-Atomtronics-MURI, and AFOSRJQI- MURI.

\bibliographystyle{apsrev}
\bibliography{sppair}

\begin{widetext}

\newpage

\begin{center}
{\Large\bf Supplementary Materials}
\end{center}

\renewcommand{\thesection}{S-\arabic{section}} \renewcommand{\theequation}{S%
\arabic{equation}} \setcounter{equation}{0} 
\renewcommand{\thefigure}{S\arabic{figure}} \setcounter{figure}{0}

\section{Hopping term} The hopping term $H_{0}$ in Eq.~\eqref{eq:Ham}
can be written as
\begin{eqnarray}
H_{0} &=&\sum_{\mathbf{R}}[C^{\dagger }(\mathbf{R})T_{0}C(\mathbf{R}%
)+C^{\dagger }(\mathbf{R})T_{1x}C(\mathbf{R+e}_{x})  \notag \\
&+&C^{\dagger }(\mathbf{R})T_{1x}^{^{\prime
}}C(\mathbf{R-e}_{x})+C^{\dagger
}(\mathbf{R})T_{1y}C(\mathbf{R+e}_{y})  \notag \\
&+&C^{\dagger }(\mathbf{R})T_{1y}^{^{\prime
}}C(\mathbf{R-e}_{y})+C^{\dagger
}(\mathbf{R})T_{2}C(\mathbf{R+e}_{x}-\mathbf{e}_{y})  \notag \\
&+&C^{\dagger }(\mathbf{R})T_{2}^{^{\prime }}C(\mathbf{R-e}_{x}+\mathbf{e}%
_{y})]  \notag \\
&&
\end{eqnarray}%
where the matrices $T$ and $T^{\prime }$ are given as
\begin{eqnarray*}
T_{0} &=&\left(
\begin{array}{ccccc}
0 & 0 & 0 & 0 & 0 \\
0 & 0 & 0 & -t_{2} & 0 \\
0 & 0 & 0 & 0 & t_{1} \\
0 & -t_{2} & 0 & 0 & 0 \\
0 & 0 & t_{1} & 0 & 0%
\end{array}%
\right), T_{2}=\left(
\begin{array}{ccccc}
0 & 0 & 0 & 0 & 0 \\
0 & 0 & 0 & -t_{2} & 0 \\
0 & 0 & 0 & 0 & t_{1} \\
0 & 0 & 0 & 0 & 0 \\
0 & 0 & 0 & 0 & 0%
\end{array}%
\right), T_{1x}=\left(
\begin{array}{ccccc}
-t_{0} & 0 & 0 & 0 & 0 \\
0 & 0 & t_{3} & t_{1} & 0 \\
0 & t_{3} & 0 & 0 & -t_{2} \\
0 & 0 & 0 & 0 & t_{3} \\
0 & 0 & 0 & t_{3} & 0%
\end{array}%
\right), \\
T_{2}^{\prime } &=&\left(
\begin{array}{ccccc}
0 & 0 & 0 & 0 & 0 \\
0 & 0 & 0 & 0 & 0 \\
0 & 0 & 0 & 0 & 0 \\
0 & -t_{2} & 0 & 0 & 0 \\
0 & 0 & t_{1} & 0 & 0%
\end{array}%
\right), T_{1y}^{^{\prime }}=\left(
\begin{array}{ccccc}
-t_{0} & 0 & 0 & 0 & 0 \\
0 & 0 & -t_{3} & t_{1} & 0 \\
0 & -t_{3} & 0 & 0 & -t_{2} \\
0 & 0 & 0 & 0 & -t_{3} \\
0 & 0 & 0 & -t_{3} & 0%
\end{array}%
\right), \\
T_{1y} &=&\left(
\begin{array}{ccccc}
-t_{0} & 0 & 0 & 0 & 0 \\
0 & 0 & -t_{3} & 0 & 0 \\
0 & -t_{3} & 0 & 0 & 0 \\
0 & t_{1} & 0 & 0 & -t_{3} \\
0 & 0 & -t_{2} & -t_{3} & 0%
\end{array}%
\right), T_{1x}^{^{\prime }}=\left(
\begin{array}{ccccc}
-t_{0} & 0 & 0 & 0 & 0 \\
0 & 0 & t_{3} & 0 & 0 \\
0 & t_{3} & 0 & 0 & 0 \\
0 & t_{1} & 0 & 0 & t_{3} \\
0 & 0 & -t_{2} & t_{3} & 0%
\end{array}%
\right)\,.
\end{eqnarray*}%
Here $t_{0}$ is the hopping amplitude between s orbital fermions;
$t_{1}$ and $t_{2}$ are the longitudinal $\sigma $-bond and
transverse $\pi$-bond hopping amplitude for p orbitals,
respectively; $t_{3}$ is the hopping
amplitude between $p_{x}$ and $p_{y}$ orbitals and $C(\mathbf{R}%
)=\left(
\begin{array}{c}
C_{s}^{A}(\mathbf{R}) \\
C_{p_{x}}^{A}(\mathbf{R}) \\
C_{p_{y}}^{A}(\mathbf{R}) \\
C_{p_{x}}^{B}(\mathbf{R}) \\
C_{p_{y}}^{B}(\mathbf{R})%
\end{array}%
\right)$ is the fermion annihilation operator located at
$\mathbf{R}=(x,y)$.

\section{Cooper's Problem}

From hopping term $H_0$ in Eq.~\eqref{eq:Ham}, we find that there
are five
Bloch bands. The corresponding operators $\alpha _s (\mathbf{k})$ and $%
\alpha _{pn} (\mathbf{k})$ for $s$ and $p$ bands are introduced,
respectively. Because the width of p band is much larger than that
of s band, intuitively we know that the condensation of Cooper pairs
between these two bands at center-of-mass momentum
$\mathbf{Q}=(\pi/a,\pi/a)$, which is the energy minimal of p band,
will be energetically favorable. Besides this intuitive picture,
systematically, this conclusion is borne out by solving energy
spectra of Cooper's bound states, which are defined as $|\Phi\rangle
= \textstyle \sum_{\mathbf{k}, \mathbf{k}^{\prime },n } ^{\prime
}\phi_n (\mathbf{k}, \mathbf{k}^{\prime }) \alpha_{pn} ^{\dag}
(\mathbf{k}) \alpha^\dag _s (\mathbf{k}^{\prime })
|\Omega \rangle,$ where $|\Omega\rangle $ is the vacuum state and $\phi_n (%
\mathbf{k}, \mathbf{k}^{\prime })$ is the two-particle wavefunction.
The summation $\sum ^{\prime }$ here is over modes above the fermi
level. Due to translational symmetry, the center-of-mass momentum
$\mathbf{Q} = \mathbf{k} + \mathbf{k}^\prime$ is a good quantum
number, which is used to label the energy spectra obtained from the
eigenvalue problem, $H |\Phi(\mathbf{Q}) \rangle = E(\mathbf{Q}) |\Phi (%
\mathbf{Q}) \rangle.$ Resulting from lattice rotation symmetry
$C_4$, there are two branches of Cooper's bound states, which are
related to each other by rotation. These two branches are most clear
in the limit of $t_3 \to 0$, i.e., without coupling between $p_x$
and $p_y$ orbitals. In this case, particle numbers of $p_x$ and
$p_y$ orbitals are separately conserved. One type of bound state is
formed by $p_x$ and $s$ orbital fermions leading to
an energy dispersion $E_x(\mathbf{Q})$; while the other formed by $p_y$ and $%
s$ orbitals leads to a dispersion $E_y(\mathbf{Q})$.

\begin{figure}[t]
\begin{center}
\includegraphics[width=8cm]{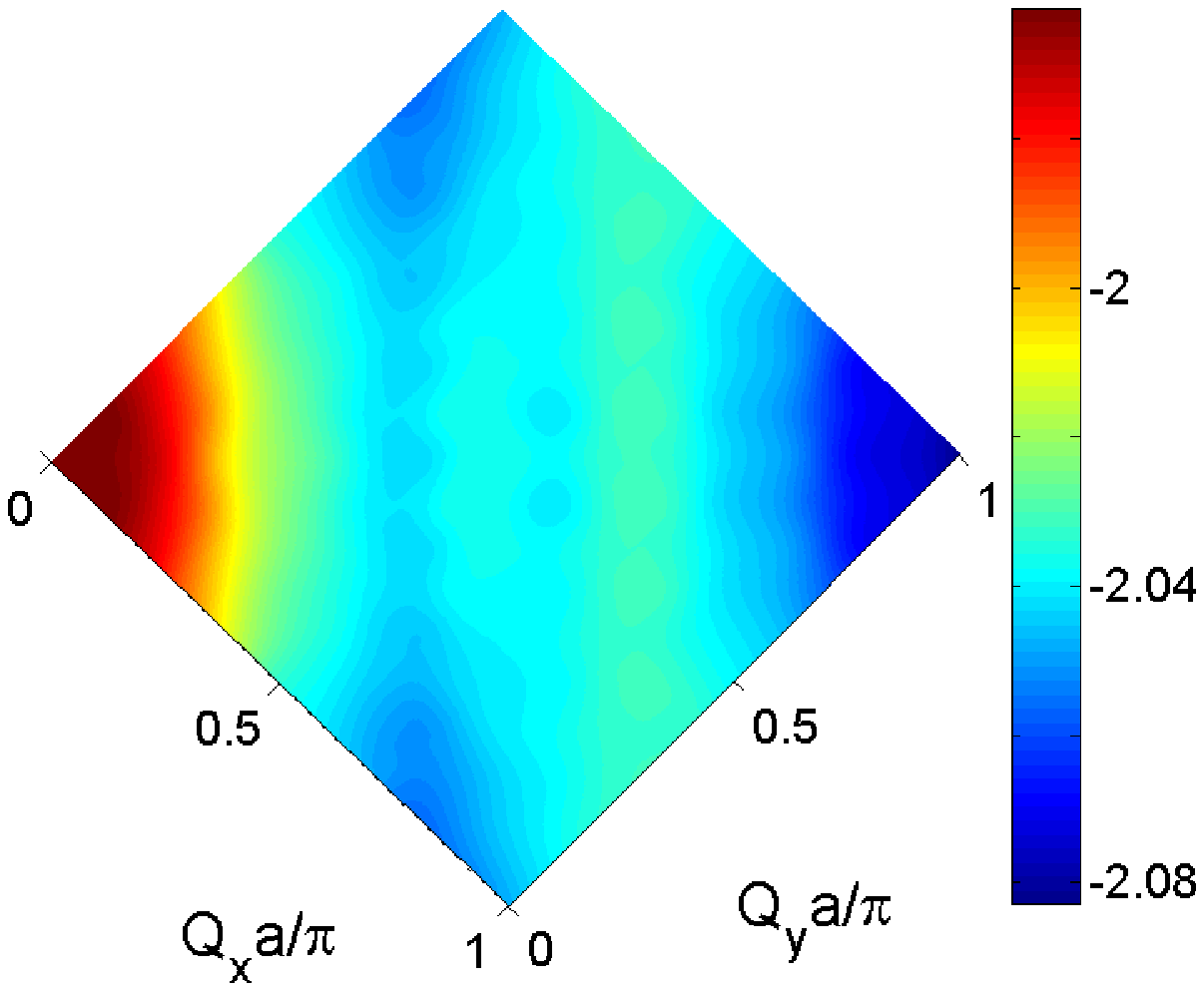}
\end{center}
\caption{Bound state energy $E_x(\mathbf{Q})/t_0$ varies as a
function
of center-of-mass momentum $\mathbf{Q}$, when $t_1/t_0=8$, $t_2/t_0=2$, $%
t_3/t_0=0$ and $U/t_0=10$.} \label{fig:bound}
\end{figure}

As shown in Fig.~\ref{fig:bound}, we find that the bound state energy $E_x(%
\mathbf{Q})$ varies as a function of center-of-mass momentum
$\mathbf{Q}$ and the energy minimal is located at
$\mathbf{Q}=(\pi/a,\pi/a)$. Due to $C_4$
symmetry, the energy minimal of $E_y(\mathbf{Q})$ also locates at $%
(\pi/a,\pi/a)$. Condensation of Cooper pairs at $\mathbf{Q}%
=(\pi/a,\pi/a)$ is energetically favorable. The effect of finite coupling $%
t_3$ between $p_x$ and $p_y$ orbitals has been discussed in the
frame work of effective field theory.

\section{Path Integral approach}

To calculate free energy from the path integral method, we introduce
the Grassman fields $\bar{\Psi }(\mathbf{R},\tau )$ and $\Psi
(\mathbf{R},\tau )$ and express the grand partition function of the
system  as
\begin{equation}
Z=\int D \bar{\Psi }D\Psi \exp (-S[\bar{\Psi },\Psi ])
\end{equation}%
with $\Psi (\mathbf{R},\tau )=\left(
\begin{array}{c}
\Psi _{s}^{A}(\mathbf{R},\tau ) \\
\bar{\Psi }_{p_{x}}^{A}(\mathbf{R},\tau ) \\
\bar{\Psi }_{p_{y}}^{A}(\mathbf{R},\tau ) \\
\bar{\Psi }_{p_{x}}^{B}(\mathbf{R},\tau ) \\
\bar{\Psi }_{p_{y}}^{B}(\mathbf{R},\tau )%
\end{array}%
\right)$. The quartic term in the interaction term of action S can
be decoupled with the Hubbard-Stranovich transformations,
\begin{align}
\tilde{\Delta} _{x}(\mathbf{R,}\tau )& =U\Psi
_{p_{x}}^{A}(\mathbf{R},\tau
)\Psi _{s}^{A}(\mathbf{R},\tau )\,,  \notag \\
\tilde{\Delta} _{y}(\mathbf{R,}\tau )& =U\Psi
_{p_{y}}^{A}(\mathbf{R},\tau )\Psi _{s}^{A}(\mathbf{R},\tau )\,.
\end{align}%
Then the partition function can be written as
\begin{equation}
Z=\int D\tilde{\Delta}_{x}D\tilde{\Delta}_{x}^{\ast }D\tilde{\Delta}_{y}D%
\tilde{\Delta}_{y}^{\ast }D\bar {\Psi }D\Psi \exp (-S[\bar {\Psi },\Psi ,%
\tilde{\Delta}_{x},\tilde{\Delta}_{x}^{\ast },\tilde{\Delta}_{y},\tilde{%
\Delta}_{y}^{\ast }]) \,. \label{eq:partition}
\end{equation}%
The action in Eq.~\eqref{eq:partition} is
\begin{eqnarray}
&&S[\bar{\Psi },\Psi ,\tilde{\Delta}_{x},\tilde{\Delta}_{x}^{\ast },\tilde{%
\Delta}_{y},\tilde{\Delta}_{y}^{\ast }]  \notag \\
&&\text{ }=\int d\tau d\mathbf{R}\{(\frac{|\tilde{\Delta} _{x}(\mathbf{R,}%
\tau )|^{2}}{U}+\frac{|\tilde{\Delta} _{y}(\mathbf{R,}\tau
)|^{2}}{U})-\int
d\tau^{\prime } d\mathbf{R'}\bar{\Psi }(\mathbf{R},\tau )G^{-1}(\mathbf{R},\tau ;\mathbf{R}%
^{\prime },\tau^{\prime })\Psi (\mathbf{R'},\tau^{\prime} )\}\,,  \notag \\
&&  \label{eq:action}
\end{eqnarray}
where $\int d\mathbf{R}=\sum_{\mathbf{R}}$. After doing an unitary
transformation of fermionic fields, we replace $\tilde{\Delta}_{x}$
and $\tilde{\Delta}_{y}$ in
Eq.~\eqref{eq:action} by two slowly varying and time-independent bosonic fields $\Delta_{x}(%
\mathbf{x})$ and $\Delta_{y}(\mathbf{x})$, respectively. Integrating
the fermionic fields, we get an effective action
\begin{equation}
S_{eff}[\Delta _{x},\Delta _{y}]=\int d\tau
d^{2}\mathbf{x}(\frac{|{\Delta} _{x}|^{2}}{U}+\frac{|{\Delta}
_{y}|^{2}}{U}-\ln\det G^{-1}[\Delta _{x},{\Delta}_{x}^{\ast },\Delta
_{y},{\Delta}_{y}^{\ast }])  \label{eq:actionmean}
\end{equation}
where $G^{-1}$ is the inverse Green's function and $\int
d^2\mathbf{x}=\sum_{\mathbf{R}}$. By calculating free energy from
Eq.~\eqref{eq:actionmean}, we obtain coefficients $r$, $g_1$, $g_2$
and $g_3 $ in Eq.~\eqref{eq:Totalfreeenergy}. As shown in
Fig.~\ref{fig:coefficients}(a), since the low temperature limit is
much smaller than the Fermi energy,  the coefficient $r$ changes
sign from positive to negative with increasing $U/t_0$, which
implies a second order phase transition from normal to a superfluid
state with $\Delta_{x\setminus y} \neq 0$ at mean field level
(Fig.~\ref{fig:phasediag}). Our numerics also find that $0<g_1 <
g_2/2$ and $|g_3| \ll g_1$. Minimizing the free energy gives a field
configuration with $|\Delta_x| = |\Delta_y|$. The relative phase
between $\Delta_x$ and $\Delta_y$ is fixed by $g_3$ as shown in
Fig.~\ref{fig:coefficients}(b) and (d). The coupling $g_3>0$ makes
the relative phase locked at $\pm \frac{\pi}{2}$ and leads to a $p_x
\pm ip_y$ superfluid state where the `$\pm$' sign is spontaneously
chosen; while $g_3<0$ favors a $p_x \pm p_y$ state
(Fig.~\ref{fig:phasediag}).

\begin{figure}[t]
\begin{center}
\includegraphics[width=8cm]{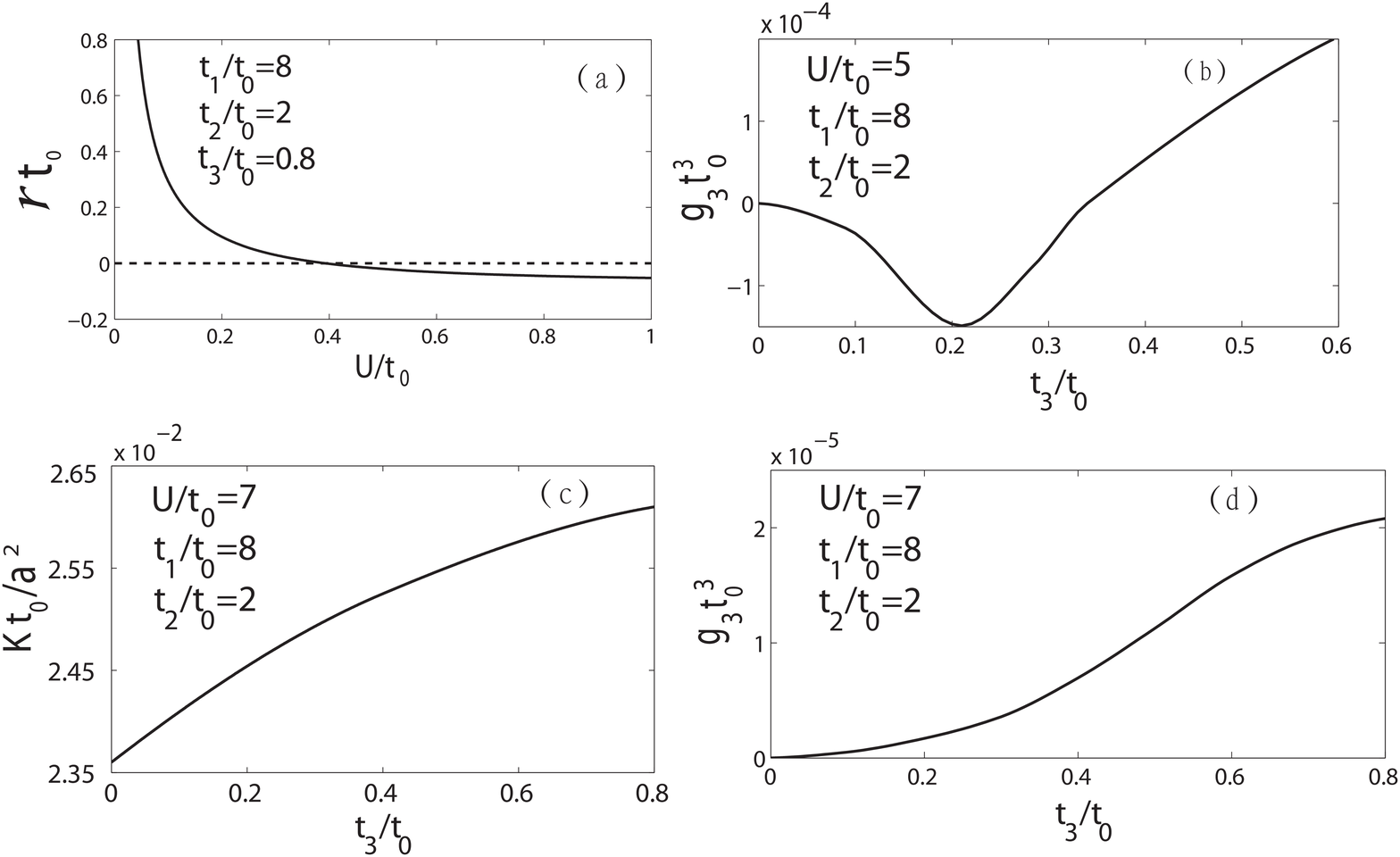}
\end{center}
\caption{(a) The coefficient $r$ vs. interaction strength $U/t_0$;
(b) and (d) The coefficient $g_3$ as a function of $t_3/t_0$; (c)
The coefficient $K$ vs. $t_3/t_0$.} \label{fig:coefficients}
\end{figure}

\section{Finite temperature phase transition}

It is well known that in 2D the transition from the normal to
superfluid state is of the Kosterlitz-Thouless type. To obtain the
KT transition temperature, we should rewrite the complex order
parameters $\Delta _{x}(\mathbf{x})=\Delta _{0}e^{i\theta
_{x}(\mathbf{x})}$ and $\Delta _{y}(\mathbf{x})=\Delta
_{0}e^{i\theta _{y}(\mathbf{x})}$ with the phase fluctuations
$\theta _{x}$ and $\theta _{y}$. Introducing new variables $\theta
=\frac{1}{2}(\theta _{x}+\theta _{y})$ and $\Delta \theta =\theta
_{x}-\theta _{y}$, from the Gaussian fluctuation part of free energy
in Eq.\eqref{eq:Totalfreeenergy}, we derive the well-known XY model
in terms of $\theta $ as $\ \Delta F[\Delta _{x},\Delta _{y}]=\int
d^{2}\mathbf{x} \tilde{K} (T) [(\partial _{x}\theta )^{2}+(\partial
_{y}\theta )^{2}]$. Here, the relative phase $\Delta \theta $ is
determined by the sign of $g_{3}$ in Eq.~\eqref{eq:Totalfreeenergy}
at finite temperature. Specifically, $\Delta \theta $ is locked at
$\pm \frac{\pi }{2}$ [or $0$] for $(p_{x}\pm ip_{y})$ [or $(p_{x}\pm
p_{y})$]. The KT transition temperature is determined by the formula
$k_{B}T_{\rm KT}={\frac{\pi}{a^2}} \tilde{K} (T = T_{\rm KT}) $.
Solving this equation self-consistently, we  get the KT transition
temperature and plot it in Fig.~\ref{fig:KTtransition}. We find that
in the weak-coupling regime $T_{\rm KT}$  approaches the mean-field
transition temperature $T_{\rm Mean}$ as determined by $r=0$ in
Eq.~\eqref{eq:Totalfreeenergy} at finite temperature. With stronger
interaction, there is a large derivation of the two as expected
~\cite{2004_Dupuis_PRB}, for the reason that mean field analysis
underestimates fluctuation effects. Our numerics also find that
$k_BT_{KT}$ can reach $3.21$$t_{0}$ accompanying with increasing of
interaction strength when the lattice strengths are $V_{s}/E_{R}=5$
and $V_{p}/2E_{R}=5$ for $s$ and $p$ orbitals, respectively. As
shown in the inset plot of Fig.~\ref{fig:KTtransition}, we also find
that decreasing of lattice strength will increase $T_{KT}$.

\begin{figure}[t]
\begin{center}
\includegraphics[width=8cm]{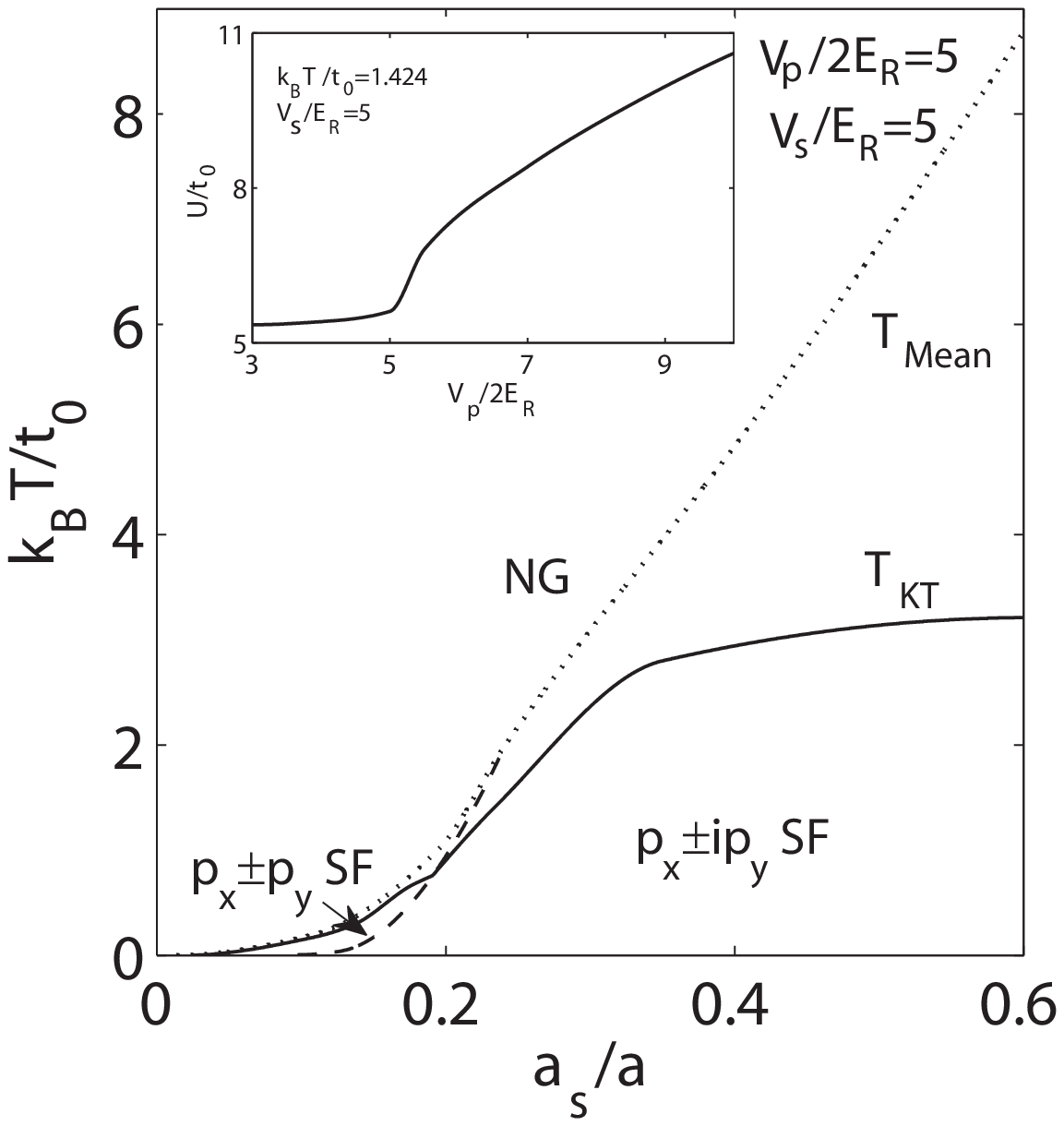}
\end{center}
\caption{Finite-temperature phase diagram--The solid line
illustrates the KT transition temperature. The mean-field transition
temperature is shown by the dot line. The regions for $p_x\pm ip_y$
and $p_x\pm p_y$ superfluid state are separated by the dash line.
Here, the lattice strengths are $V_{s}/E_{R}=5$ and $V_{p}/2E_{R}=5$
with recoil energy $E_R=\frac{h^2}{2m(2a)^2}$ and $a_s$ is the
s-wave scattering length. The inset plot shows that increasing of
lattice strength will decrease superfluid transition temperature. }
\label{fig:KTtransition}
\end{figure}

\section{Anisotropic superconducting gap}

In this section, we discuss the superconducting gap for fermions
resulting from this unconventional paring. The superconducting gap
is calculated by solving the Mean field Hamiltonian
(Eq.~\eqref{eq:Hmean}) without a domain wall defect. The anisotropy
of the gap which is a remarkable property being absent in the
conventional s-wave superconductors is characterized by
the structure functions $S_{x}(\mathbf{k}) =U/N<C_{p_{x}}^{A}(\mathbf{-k+Q}%
)C_{s}^{A}(\mathbf{k})>$ and $S_{y}(\mathbf{k}) =U/N<C_{p_{y}}^{A}(\mathbf{-k+Q%
})C_{s}^{A}(\mathbf{k})>$, where $N$ is the total site.

\begin{figure}[t]
\begin{center}
\includegraphics[width=8cm]{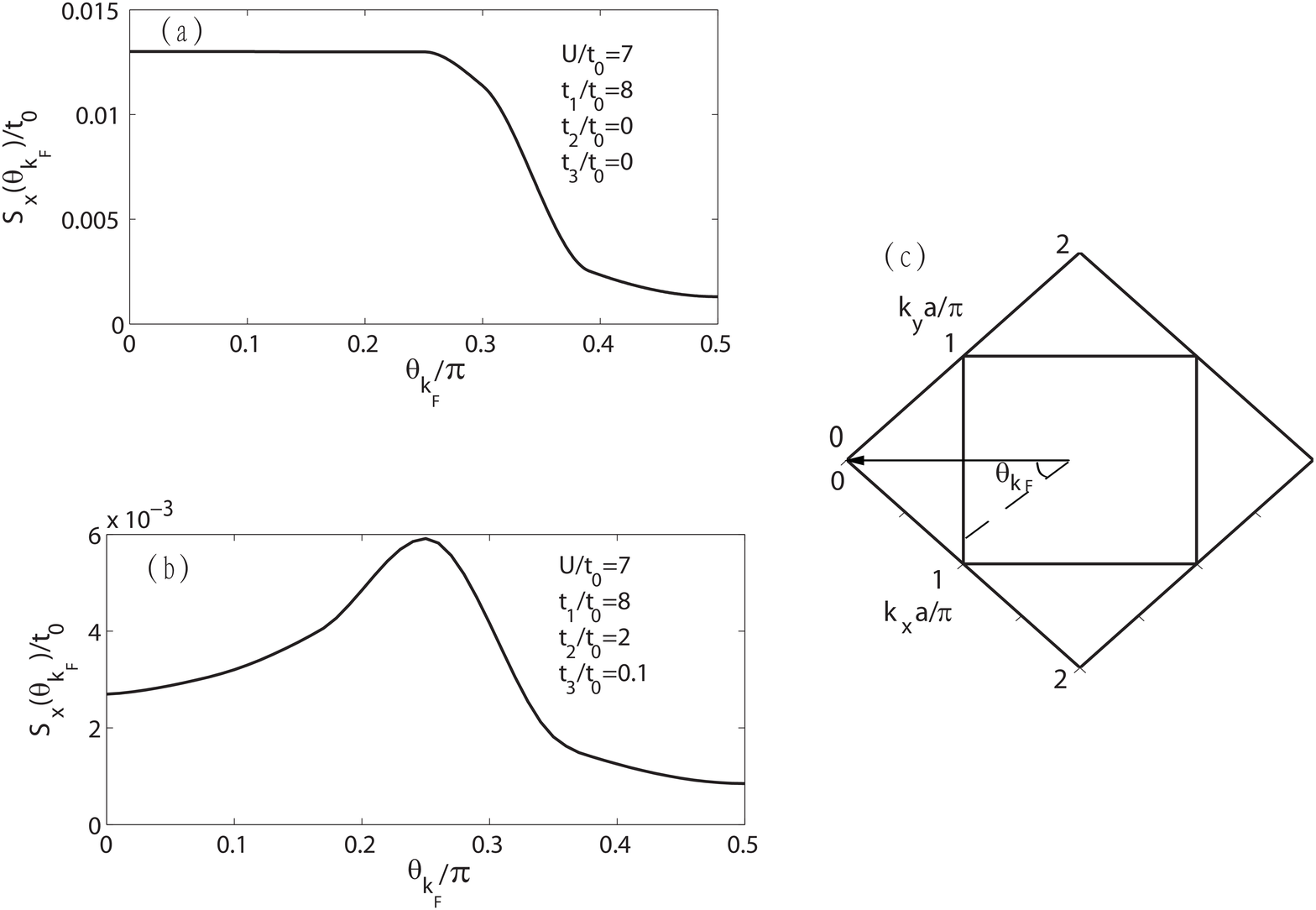}
\end{center}
\caption{Structure function $S_x (\protect\theta_{k_F})$ of
superconducting
gap near Fermi surfaces. (a) $U/t_0=7$, $t_1/t_0=8$, $t_2/t_0=0$ and $%
t_3/t_0=0$; (b) $U/t_0=7$, $t_1/t_0=8$, $t_2/t_0=2$ and $t_3/t_0=0.1$; (c) $%
\protect\theta_{k_F}$ on the Fermi surface of s orbital band. Due to
$C_4$ symmetry, the structure of $S_y$ is readily given by a
$\protect\pi/2$ rotation.} \label{fig:Sfactor}
\end{figure}

We find $S_{x/y}(\mathbf{k})$ near the Fermi surface is highly
anisotropic, that is it strongly depends on the polar angle of
$\mathbf{k}$, $\theta _{k_{F}}$, as shown in Fig.~\ref{fig:Sfactor}.
In Fig.~\ref{fig:Sfactor}(a), when $t_{2}=0$ and $t_{3}=0$, the
Fermi surface of $p$ and $s$ orbital bands fermions are matched very
well when $0\leq \theta _{k_{F}}\leq \pi /4$, so the gap are almost
at the same maximum value when $\theta _{k_{F}}$ in that region.
However, when $\pi /4<\theta _{k_{F}}<\pi /2$, the gap decreases by
increasing $\theta _{k_{F}}$ due to the mismatch of Fermi surface.
The situation is different for $t_{2}\neq 0$ and $t_{3}\neq 0$,
where the Fermi surfaces are mismatched. The gap is non-monotonic
when $\theta _{k_{F}}$
varies from $0$ to $\pi /2$, and it is maximal at $\theta =\pi /4$ (Fig.~\ref%
{fig:Sfactor}(b)). This peculiar non-monotonic behavior is related
to van-Hove singularities which lead to large density of states
nearby.

\end{widetext}


\end{document}